\documentclass[conference,a4paper]{IEEEtran}
\usepackage{cite}

\usepackage{graphicx,color,epsfig,rotating,subfigure}
\usepackage{amsfonts,amsmath,amssymb}
\usepackage{algorithm}
\usepackage{subfigure}
\usepackage{algpseudocode}

\long\def\comment#1{}

\newfont{\bbb}{msbm10 scaled 700}

\newfont{\bb}{msbm10 scaled 1100}
\newcommand{\CC}{\mbox{\bb C}}

\newcommand{\Hm}{{\bf H}}

\newcommand{\Ac}{{\cal A}}

\newcommand{\Cc}{{\cal C}}

\newcommand{\SNR}{{\sf SNR}}
\newcommand{\INR}{{\sf INR}}

\newcommand{\eqdef}{\stackrel{\Delta}{=}}

\newtheorem{theorem}{Theorem}
\newtheorem{lemma}{Lemma}

\begin{document}

\sloppy

\title{Demystifying the Scaling Laws of Dense Wireless Networks: No Linear Scaling in Practice}

\author{
  \IEEEauthorblockN{Song-Nam Hong and Giuseppe Caire}
  \IEEEauthorblockA{University of Southern California, Los Angeles, USA\\
    Email: \{songnamh,caire\}@usc.edu}
}

\maketitle

\begin{abstract}
We optimize the hierarchical cooperation protocol of Ozgur, Leveque and Tse, which is supposed to
yield almost linear scaling of the capacity of a dense wireless network with the number of users $n$.
Exploiting recent results on the optimality of
``treating interference as noise'' in Gaussian interference channels, we are able to optimize the achievable average per-link rate and not just its scaling law.
Our optimized hierarchical cooperation protocol significantly outperforms
the originally proposed scheme.
On the negative side, we show that even for very large $n$, the rate scaling is far from linear,
and the optimal number of stages $t$ is less than 4, instead of $t \rightarrow \infty$ as required for almost linear scaling.
Combining our results and the fact that, beyond a certain user density, the network  capacity is
fundamentally limited by Maxwell laws, as shown by Francheschetti, Migliore and Minero, we argue that
there is indeed no intermediate regime of linear scaling for dense networks in practice.
\end{abstract}

\section{Introduction}

Although it is extremely hard to characterize the exact capacity of wireless networks,
much progress has been made recently in the understanding of their theoretical limits.
In \cite{Ozgur}, a hierarchical protocol named {\em hierarchical cooperation}
was proposed by combining local communication and long-range cooperative MIMO communication.
Applying $t$ stages of the basic cooperative scheme to a dense network with $n$ users
in a hierarchical architecture, a capacity scaling of $\Theta(n^{\frac{t}{t+1}})$ was shown to be achievable. Therefore, for any $\epsilon > 0$,
any scaling of $\Theta(n^{1-\epsilon})$ is achievable for sufficiently large $t$. Such ``linear scaling'' of the network capacity with the number
of users $n$ is the holy grail of large wireless networks since it yields constant average rate per source-destination pair in the case where
sources and destinations are randomly selected such that their distance is $O(1)$. This, in turn, implies that the network is ``scalable'' since
the rate per end-to-end communication session does not vanish as the number of users grow. In contrast, well-known protocols such as ``decode and forward'' (aka, multi-hop  routing) yield the well known scaling of $\Theta(\sqrt{n})$ \cite{Gupta}.

While scaling law analysis yields nice and clean results, it is hard to tell how a network really performs in terms of rates, since they fail to characterize
the constants of the leading term in $n$ versus the next significant terms. Therefore, there might be significant regimes where the linear scaling does not
manifest. The purpose of this paper is twofold. First, we derive an achievable sum-rate (not just a scaling law) for the hierarchical cooperation protocol.
Second, we optimize the scheme on the basis of the achievable sum-rate, by appropriately choosing the transmit power, reuse factor, and quantization distortion level.

{\bf System model:}
We consider a network deployed over a unit-area squared region and formed by $n$ nodes placed
on a regular grid with minimum distance $1/\sqrt{n}$. The grid topology captures the essence of the problem while avoiding some
technicalities due to node random placement. The network consists of $n$ source-destination pairs, such that
each node is both a source and a destination, and pairs are selected at random over the set of $n$-permutation $\pi$
that do not fix any element (i.e., for which $\pi(i) \neq i$ for all $i = 1,\ldots, n$). We focus on max-min fairness, such that all
source-destination pairs wish to communicate at the same rate.
The channel coefficient between a transmitter node $k$ and a receive node $\ell$ at distance $r_{\ell k}$ is given by
$h_{\ell k} = r_{\ell k}^{-\alpha/2}\exp{(j\theta_{\ell k})}$, where $\alpha$ denotes the path-loss exponent
and $\theta_{\ell k}\sim\mbox{Unif}(0,2\pi]$ denotes a random i.i.d. phase.
This independent ``phase fading'' model is the same assumed in \cite{Ozgur}.

{\bf Discussion and overview of the results:}  This work gives an answer to the question of ``Is linear scaling achievable in practice?" Consider a wireless network operating on a university campus of area $\Ac=1\mbox{km}^2$. When operating around 30 GHz (i.e., $\lambda = 0.01$m), the number of spatial degrees of freedom is given by $\sqrt{\Ac}/\lambda=10^{5}$ \cite{Franceschetti}. Then, we can expect almost linear scaling up to $10^{5}$ students using hierarchical cooperation protocol in \cite{Ozgur}. However, we show that for $n \leq 10^{5}$, the optimal number of stages $t$ is less than $4$, i.e., the rate scaling is far from linear, which is in accordance with a previous result in \cite{Ghaderi}, based on the scaling law analysis, where the optimal number of stages is found to be $O(\sqrt{\log{n}})$. This apparent contradiction can be understood as follows. The linear scaling in \cite{Ozgur} is obtained by letting first $n \rightarrow \infty$ to get the scaling law of the single stage, and then $t \rightarrow \infty$ to achieve the linear scaling. In contrast, this work starts from a network density $n$ and for each $n$, we find the optimal number of hierarchical stages $t$ in terms of sum-rate, essentially capturing the impact of a finite network size.  We refer the reader to the full manuscript \cite{Hong-Long} for the detailed proofs of our results. Further, it is shown in \cite{Hong-Long}  that our optimized hierarchical cooperation scheme outperforms the classical multi-hop routing for a moderately large network size (i.e., $n\approx 10^{4}$), having a larger and larger gain as network size increases.

\section{Cooperative Transmission Scheme}\label{sec:CTS}
\begin{figure}
\centerline{\includegraphics[width=5cm]{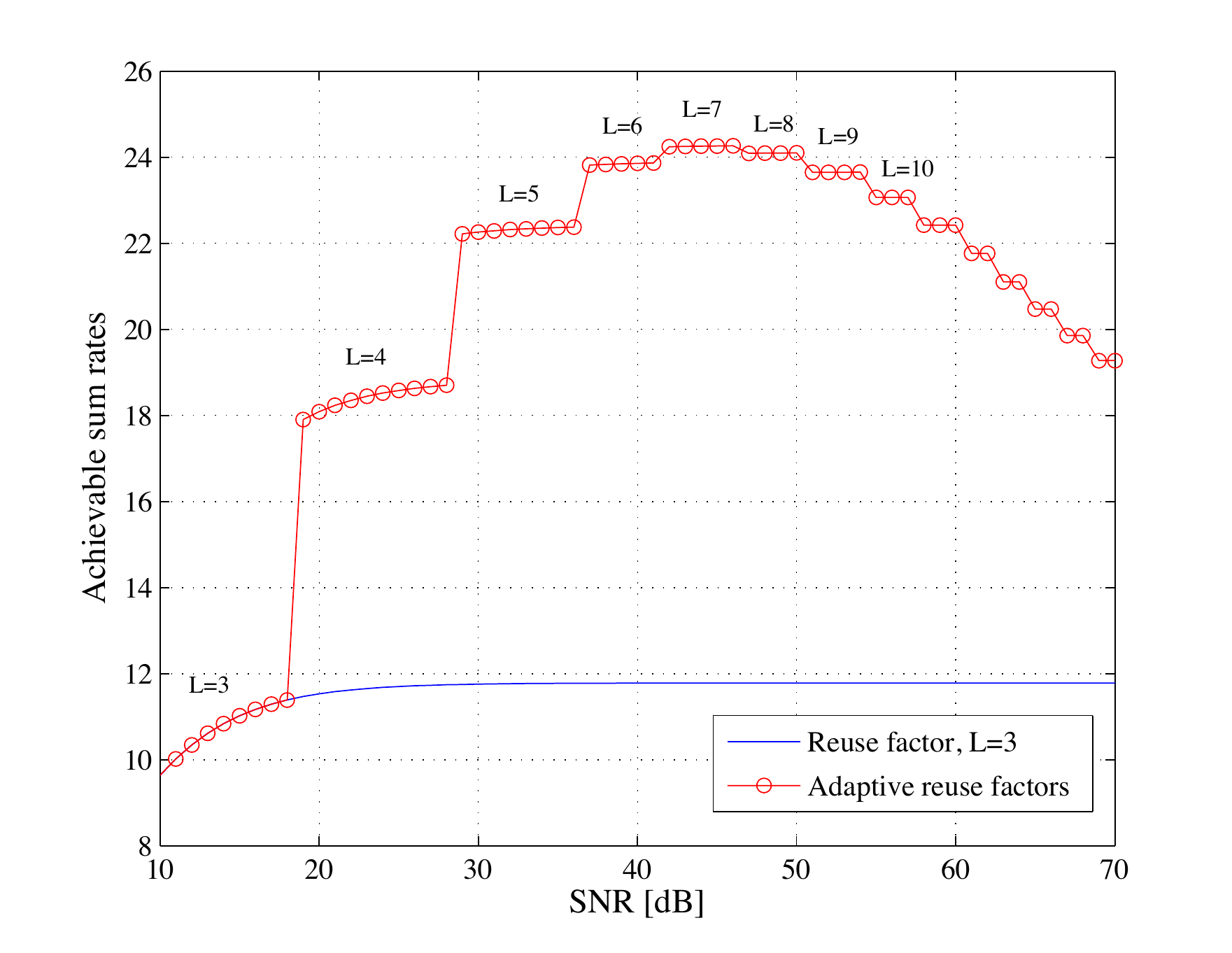}}
\caption{Achievable sum-rates of the cooperative transmission scheme when $\alpha=3$ and $n=10^{4}$. }
\label{opt_TDMA}
\end{figure}

In this section, we optimize the cooperative transmission scheme proposed in \cite{Ozgur}
with respect to the achievable sum-rate.
We let $R_{{\rm c}}(\alpha)$ denote the common message {\em coding rate} for all users, expressed in bits per codeword symbol.
The protocol delivers $n$ messages using a certain number of time-slots, each of which corresponds to the duration of a codeword.
Hence, the network sum {\em throughput} is given by $R_{\rm sum}(n,\alpha) = R_{{\rm c}}(\alpha)\mbox{T}(n,\alpha)$ where $ \mbox{T}(n,\alpha)= n/\mbox{(required number of time slots)}$
is the source-destination links per time slot ratio of the protocol (referred to as {\em packet} throughput in the following). The network is divided into $n/M$ clusters, of $M$ nodes each.
The cooperative transmission scheme consists of three phases: 1)  A ``local" communication phase is used to form cooperative clusters of transmitters.
In this phase, each source distributes $M$ distinct sub-packets of its message to the $M$ neighboring nodes in the same cluster.
One transmission is active per each cluster, in a round robin fashion, and clusters are active simultaneously
in order to achieve  some spatial spectrum reuse. The inter-cluster interference is controlled by the reuse factor $L$\footnote{All clusters have one transmission opportunity every $L^2$ time slots.};
2) A ``global'' cooperative MIMO transmission phase is used to deliver messages across different clusters. In this phase,
one cluster at a time is active, and when a cluster is active it operates as a single $M$-antenna MIMO transmitter, sending $M$ independently encoded
data streams to a destination cluster. Each node in the cooperative receiving cluster stores its own received signal;
3) A ``local'' communication phase during which all receivers in each cluster share their own received and quantized signals in order to allow each destination in the cluster to decode its intended message on the basis of the (quantized) $M$-dimensional observation. Quantization and binning (or random hashing
of the quantization bits onto channel codewords) is used in this phase, which is a special case of the general Quantize reMap and Forward (QMF) scheme
for wireless relay networks \cite{Avestimehr}. Each destination performs joint typical decoding to obtain its own desired message
based on the quantized signals (or bin indices).

The parameters we need to optimize in the cooperative transmission scheme are the cluster size $M$,
the node transmit power $\SNR$, reuse factor $L$, and quantization distortion level. Regarding the transmit power, it is assumed that $\SNR$ can be chosen arbitrarily with a uniform
bound $\SNR \leq \SNR_{\max}$ where the latter is a fixed arbitrarily large constant that does not scale with $n$. As the result of such optimization in Sections~\ref{subsec:LC}, \ref{subsec:MIMO}, and~\ref{subsec:ST1}, we have:

\begin{theorem}\label{thm:sum-rate} For any network size $n$ and path-loss exponent $\alpha$, the cooperative transmission scheme achieves the sum-rate of
\begin{equation*}
R_{\rm sum}(n,\alpha) = \log\left(1+\frac{\SNR}{1+P_{I}}\right)\frac{\sqrt{n}}{2\sqrt{2}L(\SNR)} \label{eq:sum1}
\end{equation*}where $\SNR = 2^{2(3+\alpha/\ln{2})}$, $L(\SNR) = \left\lceil\sqrt{\SNR}^{1/\alpha}+1\right\rceil$, and  $P_{I} =\sum_{i=1}^{\sqrt{n}} 8i\SNR (L(\SNR)-1)^{-\alpha}$.\hfill $\QED$
\end{theorem}


Theorem \ref{thm:sum-rate} implies that all sources can reliably transmit their messages at rate
$R_{{\rm c}}(\alpha) \approx\log{\sqrt{\SNR}}$ over the $2\sqrt{2}L(\SNR)\sqrt{n}$ time slots. Notice that despite the fact we let $\SNR_{\max}$ to be an arbitrarily large constant, the optimal SNR depends only on the pathloss $\alpha$ and it is generally not too large. This is because there is a tension between
the transmit power of each local link and the reuse factor necessary to keep inter-cluster interference under control.
The optimal transmit power is determined in Section~\ref{subsec:ST1} (Theorem~\ref{thm:sum-rate}), as a result of this tradeoff.
For comparison, notice that in the original scheme of \cite{Ozgur} the reuse factor is fixed to 3.
Fig.~\ref{opt_TDMA} shows that our  optimized scheme provides a substantial gain over the conventional scheme \cite{Ozgur}.


\begin{figure}
\centerline{\includegraphics[width=4.5cm]{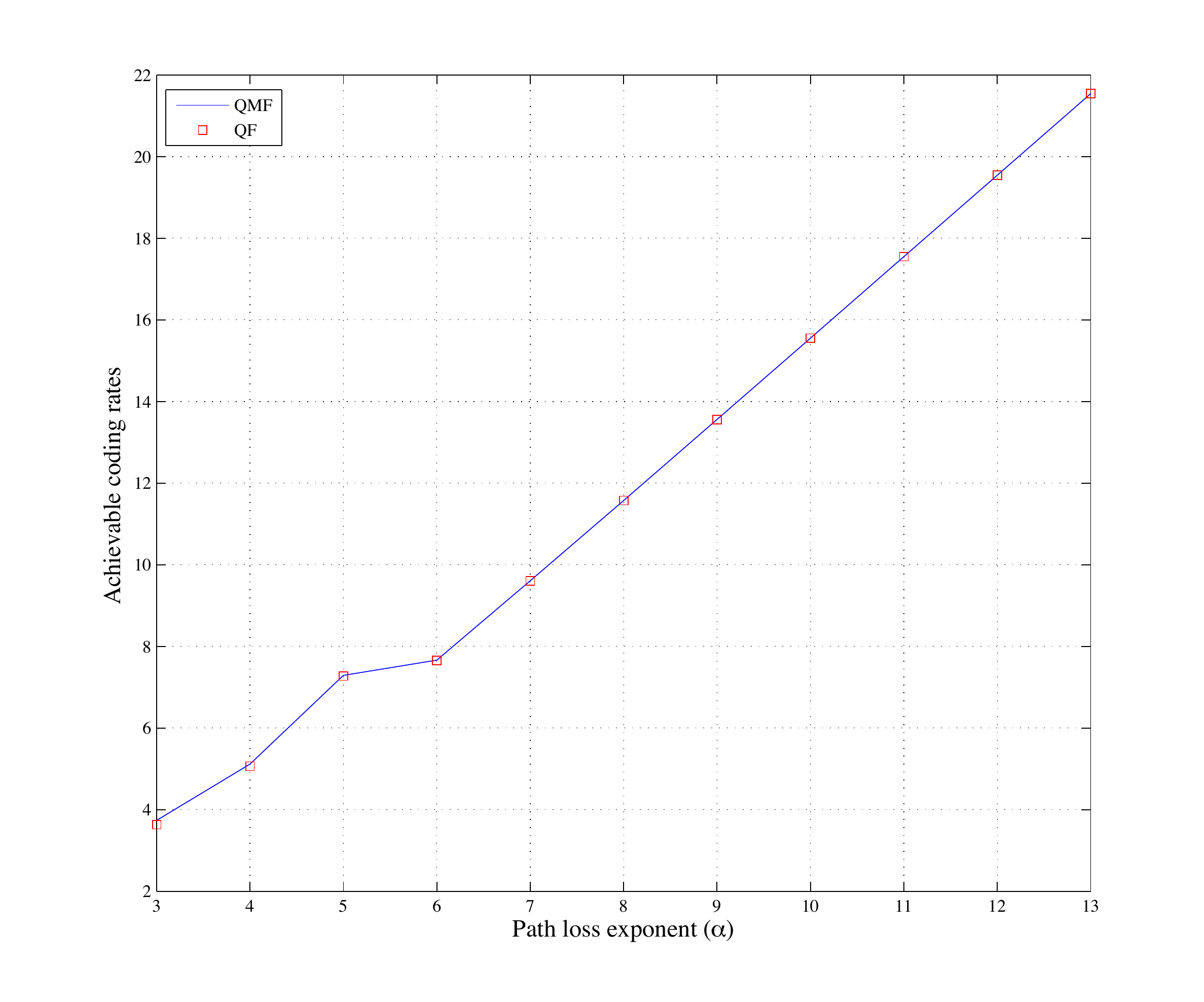}}
\caption{Achievable coding rates of hierarchical cooperation protocol as a function of path-loss exponent $\alpha$.}
\label{MRates}
\end{figure}


\subsection{Local Communication}\label{subsec:LC}

In phases 1 and 3 of the scheme there is no {\em intra-cluster interference} since a single transmitter-receiver pair is
active in each cluster.  However, each receiver suffers from inter-custer interference from the transmitters in the other clusters.
Hence, we are in the presence of a $n/M$ user (symmetric) interference channel.
It has been recently recognized that there exists a regime of the interference channel gains for which
{\em treating interference as noise} (TIN) is information-theoretically optimal (to within a constant gap) \cite[Theorem 4]{Geng}.
Furthermore, TIN is most attractive in practice since it requires standard ``Gaussian'' codes and minimum distance decoders.
Hence, we shall operate the interference channel induced by the local communication phases in the regime for which TIN is (near) optimal.
This can be obtained by choosing a reuse factor $L$ such that the TIN optimality condition on the channel coefficients of the simultaneously active
links is satisfied.
We first determine the transmit power $P$ according to the cluster area $\Ac$ and the path-loss exponent $\alpha$
as $P=\SNR\Ac^{\alpha/2}$. This choice makes that $\SNR_{i} = \SNR$ for all $i$  where $\SNR_{i}$ denotes the received power
of desired signal at node $i$. Also, let $\INR$ denote the strongest interference power, i.e., $\INR = \max_{j\neq i}\INR_{ij} = \max_{k \neq i}\INR_{ki}$,
where $\INR_{ij}$ denote the interference-to-noise ratio of source $j$ at destination $i$ and the last equality is due to the symmetric structure of network.
Considering the TDMA structure, we obtain that $\INR = (L-1)^{-\alpha}\SNR$. In our symmetric model, the optimality condition of TIN \cite{Geng}
is satisfied if $\INR \leq \sqrt{\SNR}$. We can find $L$ to meet the above condition as $L(\SNR) =  \left\lceil\sqrt{\SNR}^{1/\alpha}+1\right\rceil$.
Then, the local communication rate of  $R^{(1)}(\SNR) = \log\left(1+\SNR/(1+P_{I})\right)$ is achievable by TIN, where
the inter-cluster total interference is upper bounded by $P_{I}$ defined in Theorem 1.  
Reliable local communication is ensured by letting:
\begin{equation}
R_{{\rm c}}(\alpha) \leq R^{(1)}(\SNR). \label{eq:rate-const1}
\end{equation}

\subsection{Long-Range MIMO Communication}\label{subsec:MIMO}

Concatenating phases 2 and 3 is analogous to a distributed MIMO channel with finite backhaul capacity
of rate $R_{0}$  \cite{Sanderovich,Hong},
where the $M$ transmit (resp., $M$ receiver) antennas correspond to the $M$ nodes in the source cluster (resp., destination cluster).
For the MIMO transmission (i.e., phase 2), the transmit power is given by $P_{{\rm MIMO}}=(\SNR'/M)\Ac^{\alpha/2}$ where $\SNR'$ can be arbitrary chosen with $\SNR'\leq \SNR_{\max}$ as before.
Including the impact of distance-dependent power control into a channel, the channel matrix of distributed MIMO channel is
$\Hm \in \CC^{M \times M}$, with $(k,\ell)$-element given by $\exp(j\theta_{k\ell})$  with $\theta_{k\ell} \sim \mbox{Unif}(0,2\pi]$.
Let $N_{0}$ denote the variance of additive noise plus inter-cluster interference.\footnote{Inter-cluster interference is zero in a single layer of the
hierarchical cooperation, but is non-zero with multiple stages so it is treated in general here.}
As in \cite{Ozgur}, the local communication of phase 3 can be expanded over $Q$ time slots for some integer $Q$, in order to obtain more flexibility in the
quantization rate of the underlying QMF scheme.  This yields the backhaul capacity of the ``equivalent'' model
as $R_{0} = QR^{(1)}(\SNR)$.  An optimal $Q$ will be chosen later on.

The computation of the rate achievable by QMF for the distributed MIMO channel with finite backhaul capacity is
generally difficult  since it involves a complicated combinatorial optimization  \cite{Sanderovich}.
So far, a closed-form expression was only available for the symmetric Wyner model \cite{Sanderovich}.
In this paper, we derive a closed-form expression of  the achievable rate for our model, exploiting the fact that, for large $n$,
the problem symmetries although the matrix $\Hm$ is ``full'' and not tri-diagonal as in the Wyner model. Our result is based on
asymptotic Random Matrix Theory and the submodular structure of rate expression:

\begin{theorem}\label{thm:DASrate} For a distributed MIMO channel with backhaul capacity of $R_{0}$ and random i.i.d. channel coefficients with zero mean
and unit variance,
QMF achieves the symmetric rate of
\begin{eqnarray}
&& R_{{\rm QMF}}(R_{0},N_{0},\SNR)\label{eq:QMF}\\
&=& \min\left\{R_{0}-\log\left(1+N_{0}/\sigma_{q}^2\right), \Cc\left(\SNR/(N_{0}+\sigma_{q}^2)\right)\right\} \nonumber
\end{eqnarray} for some quantization level $\sigma_{q}^2 \geq 0$, where $\Cc(x)= 2\log\left((1+\sqrt{1+4x})/2\right)-(\sqrt{1+4x}-1)^2\log{e}/4x$. \hfill $\QED$
\end{theorem}

Since the first rate-constraint is an increasing function of $\sigma_{q}^2$ and the second rate-constraint is a decreasing function
of $\sigma_{q}^2$, the optimal value of $\sigma_{q}^2$ is attained by solving $R_{0}-\log\left(1+N_{0}/\sigma_{q}^2\right)= \Cc\left(\SNR/(N_{0}+\sigma_{q}^2)\right)$. Define $h(\sigma_{q}^2) \eqdef R_{0}-\log\left(1+N_{0}/\sigma_{q}^2\right) - \Cc\left(\SNR/(N_{0}+\sigma_{q}^2)\right)$. Then, we can find  $\sigma_{q,{\rm min}}^2 = N_{0}/ (2^{R_{0}} -1)$ and $\sigma_{q,{\rm max}}^2=(N_{0}+\SNR)/(2^{R_{0}}-1)$ such that $h(\sigma_{q,{\rm min}}^2) \leq 0$  and  $h(\sigma_{q,{\rm max}}^2) \geq 0$. This is because $\sigma^2_{q,{\rm min}}$ makes the first rate-constraint in (\ref{eq:QMF}) zero and $\sigma^2_{q,{\rm max}}$ is the
quantization level of Quantize and Forward (QF)\footnote{QF is a simplified version of QMF without using binning (see \cite{Hong} for details).}, which makes the second rate-constraint in (\ref{eq:QMF}) active. Using bisection method, we can quickly find an optimal quantization level $\sigma^2_{q,{\rm opt}}$
that will be used in this paper to plot the achievable rates of QMF.

Putting together the MIMO rate constraint of Theorem \ref{thm:DASrate} with the rate achievable in phase 1 (\ref{eq:rate-const1}), we find $R_{{\rm c}}(\alpha) = \min\{R^{(1)}(\SNR), R_{{\rm QMF}}(QR^{(1)}(\SNR),1,\SNR'))\}$.
Since there is no inter-cluster interference (i.e., $N_{0}=1$) in the MIMO communication phase 2,
we can find some finite value $\SNR'$ with $Q=1$ such that $R^{(1)}(\SNR) \leq R_{{\rm QMF}}(R^{(1)}(\SNR),1,\SNR')$.
Then, we have that $R_{{\rm c}}(\alpha) = R^{(1)}(\SNR)$, where an optimal $\SNR$ will be determined in the next section.
In fact, we do not have to compute an exact achievable rate of QMF in this section but
QMF rates will be used in Section~\ref{sec:H-CTS} for the hierarchical cooperation protocol,
when we shall consider multiple stages of the 3-phase cooperative scheme.

\subsection{Achievable sum-rate}\label{subsec:ST1}

In order to derive an achievable sum-rate, we will compute the packet throughput $\mbox{T}(n,\alpha)$.
As anticipated before, in the cooperative scheme each source transmits $M$ distinct sub-packets of the message to the intended destination.
To transmit overall $nM$ sub-packets (in the whole network), phase 1 requires the
$(L(\SNR)M)^2$ time slots, phase 2 requires $n$ time slots, and phase 3 requires the $Q(L(\SNR)M)^2$ time slots.
Based on this, we have $\mbox{T}(n,\alpha) = Mn/((Q+1)(L(\SNR)M)^2+n)$.
Since the coding rate $R^{(1)}(\SNR)$ is independent of $M$, we can find the optimal cluster size $M$
by treating $M$ as a continuous variable and solving $d \mbox{T}(n,\alpha)/ d M = 0$. This yields
$M = \sqrt{n}/(L(\SNR)\sqrt{1+Q})$. Then, the packet throughput is obtained as $\mbox{T}(n,\alpha) = \sqrt{n} / (2L(\SNR)\sqrt{1+Q})$ and accordingly, the achievable sum-rate is given by $R_{{\rm sum}} (n,\alpha) = R^{(1)}(\SNR)\sqrt{n}/(2L(\SNR)\sqrt{1+Q})$.
Next, we will optimize the transmit power $\SNR$ to maximize the above sum-rate.
To make the problem tractable, we use the approximations $L(\SNR) = \sqrt{\SNR}^{1/\alpha}$ and $R^{(1)}(\SNR) = \log(\sqrt{\SNR}/8)$.
Then, the sum-rate is approximated by $\tilde{R}_{{\rm sum}} (n,\alpha) =\sqrt{n}\log(\sqrt{\SNR}/8) /(2\sqrt{2}\sqrt{\SNR}^{1/\alpha})$
where  $Q=1$ is chosen because of the explanation given before.
Differentiating and solving $d \tilde{R}_{{\rm sum}} (n,\alpha)/d \SNR=0$, we find that the optimal transmit power is given by $\SNR= 2^{2(3+\alpha/\ln2)}$.

\begin{figure}
\centerline{\includegraphics[width=5cm]{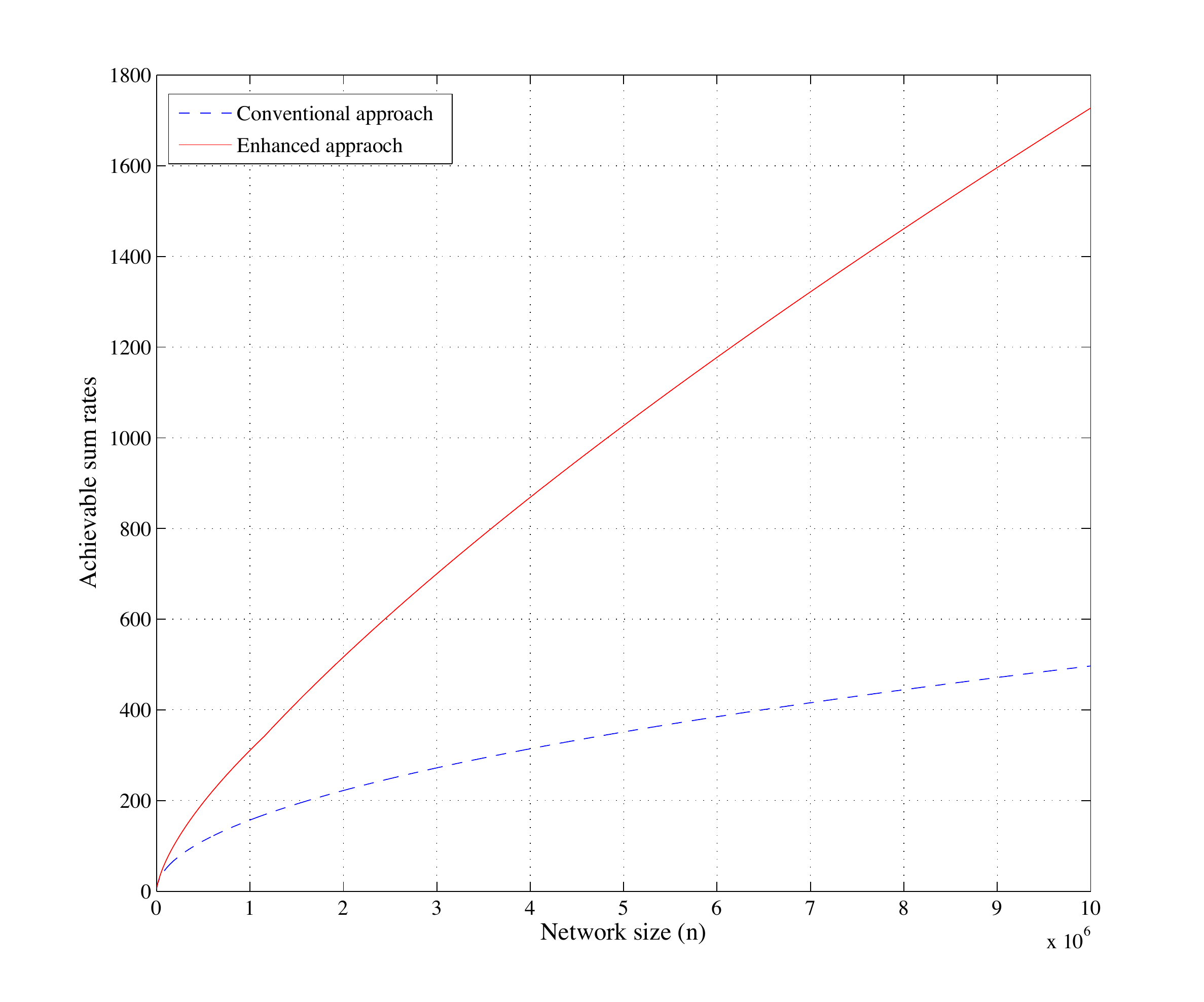}}
\caption{$\alpha=3$. Achievable sum-rates as a function of $n$.}
\label{sumrate}
\end{figure}

\section{Optimizing the hierarchical cooperation protocol}\label{sec:H-CTS}

{\em Hierarchical} cooperation was proposed in \cite{Ozgur} by employing the cooperative transmission scheme of
Section~\ref{sec:CTS} as the local communication of a higher stage. In this scheme, we use the symmetric coding
rate $R_{{\rm c}}(\alpha)$ regardless of the number of hierarchical stages $t$. Based on Section~\ref{sec:CTS}, we choose
$P_{i}= \SNR \Ac_{i}^{\alpha/2}$, $L = \left\lceil \sqrt{\SNR}^{1/\alpha} +1 \right \rceil$, and $\SNR=2^{2(3+\alpha/\ln{2})}$,
for stages $i=1,\ldots,t$, where $\Ac_{i}$ denotes the cluster area of stage $i$.
Notice that these choices guarantee that, regardless of hierarchical stage $i$, the received power of inter-cluster interference is at most
equal to $P_{I}$ in Theorem 1. For the MIMO communication phase, we choose
transmit power $P_{\mbox{{\tiny MIMO}},i}=(\SNR/M) \Ac_{i}^{\alpha/2}$, which also makes the interference power to be not larger than
$P_{I}$. The following is the main result of this section:
\begin{theorem}\label{thm:H-sumrate}
For any network size $n$ and path-loss exponent $\alpha$, the hierarchical cooperation protocol
with $t\geq 2$ stages achieves the sum-rate of
\begin{equation*}
R_{\rm sum}^{(t)}(n,\alpha) = R_{{\rm c}}(\alpha) n^{\frac{t}{t+1}}/\left((1+t)L^{\frac{2t}{t+1}}\sqrt{3}^{t}\right)
\end{equation*} where $L= \left\lceil 2^{(3+\alpha/\ln{2})/\alpha} +1 \right\rceil$ and $R_{{\tiny c}}(\alpha)$ is determined in Section~\ref{subsec:M-RATE}. Some coding rates $R_{{\rm c}}(\alpha)$ are provided for the interesting $\alpha$'s in Fig.~\ref{MRates}.
\end{theorem}
\begin{IEEEproof}
See Sections~\ref{subsec:M-RATE} and~\ref{subsec:ST}.
\end{IEEEproof}


When $t=1$, the sum-rate in the above does not reduce to the previous result in Theorem~\ref{thm:sum-rate} since in this case
we can choose a higher coding rate than $R_{{\tiny c}}(\alpha)$ in Fig.~\ref{MRates}, because there is no inter-cluster interference in the
MIMO communication phase.
From Theorem~\ref{thm:H-sumrate}, we observe that a linear scaling can be achieved as $t \rightarrow \infty$ when the network
size $n$ grows faster than the constant term $(1+t)L^{\frac{2t}{t+1}}\sqrt{3}^{t}$. However, for a finite network size,
the constant term cannot be neglected since it also grows with $t$.
Namely, adding more stages does not necessarily improve the achievable sum-rate.
Thus, for given $n$, we can find an optimal number of hierarchical stages to maximize the sum-rate.
In order to make the problem manageable, we relax the integer constraint on $t$ and find the optimal $t$ as solution of $d R_{{\rm sum}}^{(t)} (n,\alpha)/ d t = 0$.
This gives the equation in $t$ as $(t+1)^2\ln{\sqrt{3}} + (t+1) - \ln(n/L) = 0$, which yields $t_{{\rm opt}} = -1+(-1+\sqrt{1+2\ln(n/L)\ln{3}})/\ln{3}$.
This shows the following negative result: even for $n$ as large as $10^{7}$, the optimal number of hierarchical stages
is not larger than 4. Hence, for networks of reasonable size, the linear scaling law is a ``myth'', even without considering the
physical propagation limitations analyzed in \cite{Franceschetti}.
Fig.~\ref{sumrate} plots the achievable sum-rate of the hierarchical cooperation protocol
with the optimal number of stages. The conventional scheme is the one presented in \cite{Ozgur} and the enhanced scheme is the one presented in this paper
with sum-rate in Theorem~\ref{thm:H-sumrate}. In our scheme, we have modified the TDMA phases  in order to
reduce the transmission overhead (see Section~\ref{subsec:ST} for details).
We observe that the enhanced scheme provides a considerable gain over the conventional scheme,
having a larger gap as $n$ increases. Nevertheless, the network throughput is clearly sub-linear even for the range of unreasonably large $n$
shown in the figure.

\begin{figure}
\centerline{\includegraphics[width=5cm]{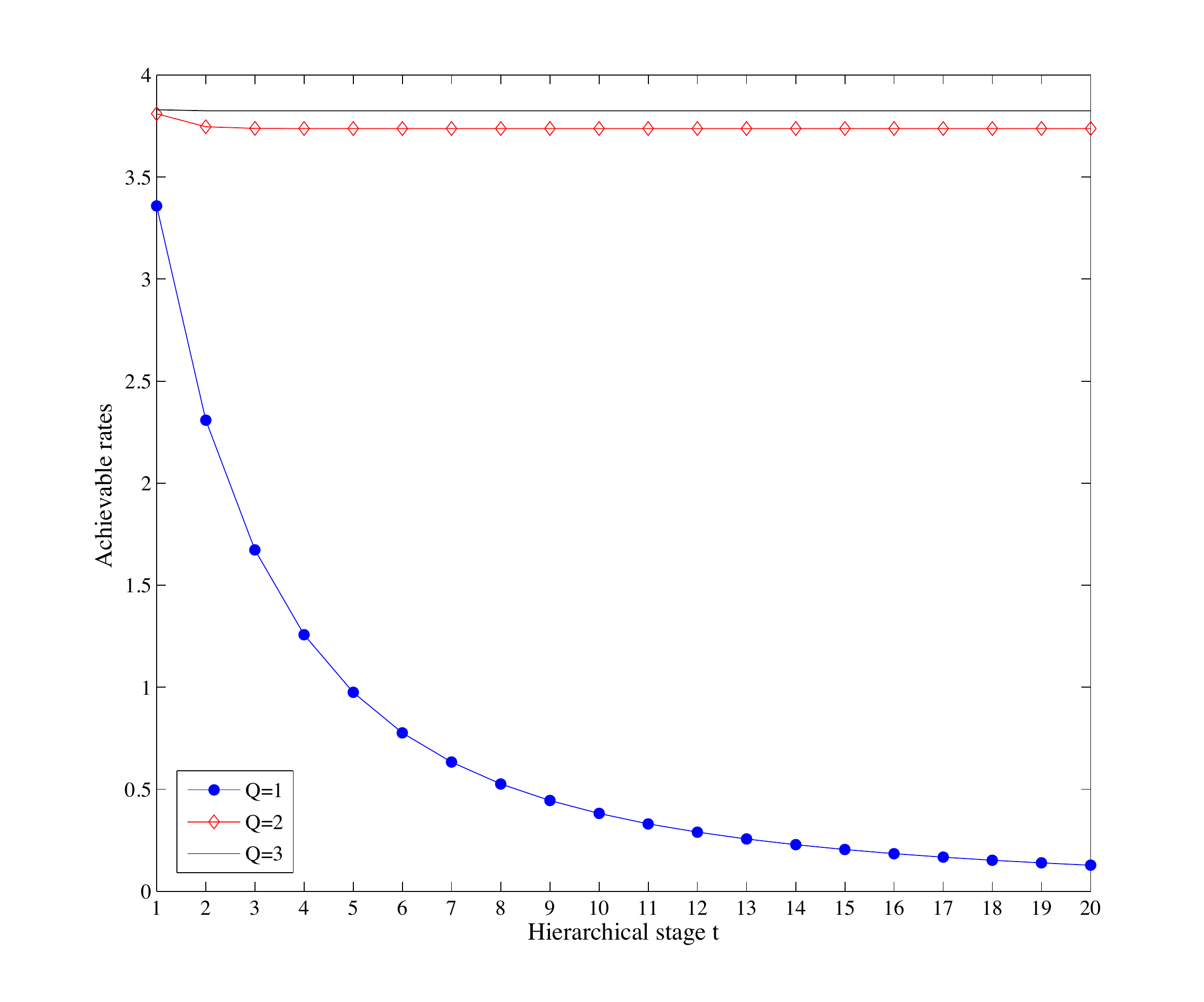}}
\caption{$\alpha=3$. Achievable coding rate  as a function of hierarchical stage $t$.}
\label{MRate}
\end{figure}

\subsection{Achievable coding rate}\label{subsec:M-RATE}

From Section~\ref{sec:CTS}, we have the rate-constraint of $R_{{\rm c}}(\alpha) \leq R^{(1)} \eqdef \log\left(1+\SNR/(1+P_{I})\right)$ for
reliable local communication at the bottom stage (i.e., stage 1).
Concatenating the phases 2 and 3 of stage 1, we can produce a distributed MIMO channel with backhaul capacity
of $QR^{(1)}$ (see Section~\ref{subsec:MIMO}). Then, the coding rate should satisfy the $R_{{\rm c}}(\alpha) \leq R_{{\rm QMF}}(QR^{(1)},N_{0}=P_{I}+1,\SNR)  \eqdef R^{(2)}$. Lemma~\ref{lem:bound} below yields that $R^{(2)} \leq R^{(1)}$ for any positive integer $Q\geq 1$.
Since $R^{(2)}$ is the local communication rate of stage 2, we can produce a {\em degraded} distributed MIMO channel with backhaul
capacity $QR^{(2)} \leq QR^{(1)}$, resulting in the rate-constraint
$R_{{\rm c}}(\alpha) \leq R_{{\rm QMF}}(QR^{(2)},N_{0}=P_{I}+1,\SNR)  \eqdef R^{(3)}$.
Clearly, we have that $R^{(3)} \leq R^{(2)}$.
Repeating the above procedures, we obtain that $R^{(t+1)} \eqdef  R_{{\rm QMF}}(QR^{(t)},P_{I}+1,\SNR) \leq R^{(t)}$, such that
$\{R^{(t)}\}$ is monotonically non-increasing. Hence, there exists a limit $\lim_{t \rightarrow \infty } R^{(t)} = R^{\star}(\alpha,Q)$, where
such limit  depends on $\alpha$ and $Q$. All rate-constraints are satisfied by choosing $R_{{\rm c}}(\alpha) = R^{\star}(\alpha,Q)$.
One might have a concern that this choice is not a good one for small $t$.  However, Fig.~\ref{MRate} shows that $R^{(t)}$ quickly converges
to its positive limit for $Q\geq 2$. Also, we observe that $Q=2$ is the best choice since it almost achieves the upper bound $R^{(1)}$,
by minimizing the required number of time slots.
Therefore, we choose the $Q=2$ and $R_{{\rm c}}(\alpha) = R^{\star}(\alpha,2)$ in the following, for any $t$.
The corresponding coding rates are plotted in Fig.~\ref{MRates} as a function of $\alpha$.

\begin{lemma}\label{lem:bound} For any $Q \geq 1$, the achievable rate of MIMO transmission is upper-bounded by the local communication rate of bottom stage (i.e., stage 1) as $R_{{\rm QMF}}(Q R^{(1)}, 1+P_{I}, \SNR) \leq R^{(1)}$.\hfill $\QED$
\end{lemma}

\subsection{Achievable sum-rate}\label{subsec:ST}

Focusing on the packet throughput, we first review the work in \cite{Ozgur} and then improve it by efficiently
using the TDMA scheme during the local communication phases.

\begin{figure}
\centerline{\includegraphics[width=6cm]{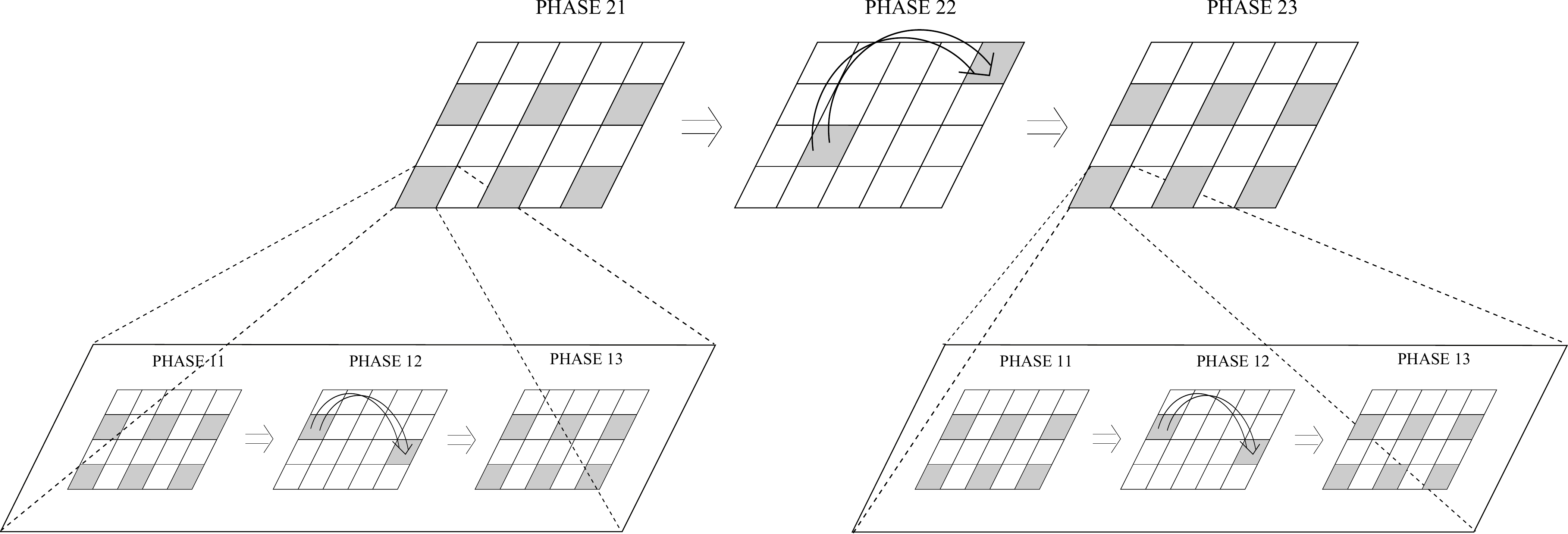}}
\caption{The silent features of the conventional scheme \cite{Ozgur} for $t=2$.}
\label{eTDMA}
\end{figure}

{\bf Approach in \cite{Ozgur}:}
The operation of stage 1 is equivalent to the cooperative transmissions and thus, from Section~\ref{subsec:ST1},
the packet throughput is computed as $\mbox{T}^{(1)}(n,\alpha) = \sqrt{n}/(2L\sqrt{1+Q})$.
Also, stage 2 employs stage 1 as its local communication (see Fig.~\ref{eTDMA}).
Then, the required number of time slots is $(LM_{1})^2 / \mbox{T}^{(1)}(M_{1},\alpha)$ for phase 1, $n$ for phase 2, and
$Q(LM_{1})^2/\mbox{T}^{(1)}(M_{1},\alpha)$ for phase 3. With the optimal cluster size $M_{1} =n^{2/3}/(L^2(1+Q))$, the resulting packet throughput is given by $\mbox{T}^{(2)}(n,\alpha) =n^{2/3}/(3(L\sqrt{1+Q})^2)$. Generalizing to $t$ stages, we obtain
the achievable sum-rate of the scheme in \cite{Ozgur} as $R_{{\rm sum}}^{(t)} (n,\alpha) = R_{{\rm c}}(\alpha)n^{t/(t+1)}/((t+1)(L\sqrt{1+Q})^{t})$.

\begin{figure}
\centerline{\includegraphics[width=6cm]{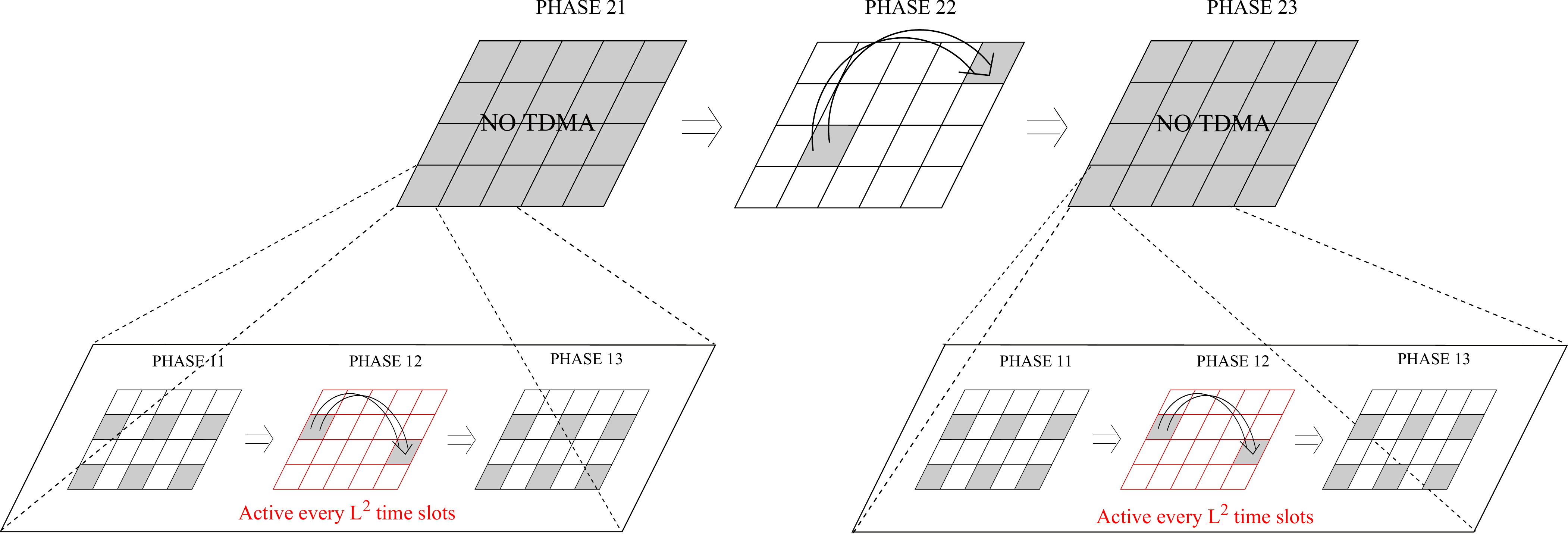}}
\caption{The silent features of the enhanced scheme for $t=2$.}
\label{eTDMA1}
\end{figure}

{\bf Enhanced approach:}
We will improve the penalty term associated with TDMA from $L^{t}$ to $L^{\frac{2t}{t+1}}$.
This provides a non-trivial gain especially when $t$ is large, since the former exponentially increases with $t$
while the latter is upper bounded by $L^2$.
We explain our approach based on a 2-stage hierarchical cooperation protocol (see Fig.~\ref{eTDMA1}) and then
extend the result to general $t$. First, we want to emphasize that TDMA scheme is used so that the received power
of interference is less than a certain level for all transmissions. It can be noticed that this requirement is satisfied
for the transmissions of phase 11 (or phase 13) (i.e., stage 1, phases 1 and 3)
without using the TDMA scheme of phase 21 (stage 2, phase 1) since local communications have already
included the TDMA operation. However, the TDMA scheme of phase 21 is required for the long-range MIMO communication of phase 12.
Based on this observation, we present an alternative approach to efficiently use the TDMA scheme  (see Fig.~\ref{eTDMA1}):
All clusters in phase 21 (or phase 23) are always active (spatial reuse 1);
In phase 12, each cluster has a turn to perform the MIMO transmissions every $L^2$ time slots,
which is equivalent to apply the TDMA scheme of phase 21 (or phase 23). In short, this approach applies the TDMA scheme only once to
every phase. Then, we can recompute the required number of time slots as follows.
Since TDMA scheme is used for all phases in stage 1, it requires the $(LM_{1})^2+L^2n+Q(LM_{1})^2$ time slots.
With the optimal cluster size $M_{1} = \sqrt{M_{2}}/\sqrt{1+Q}$, we can compute the packet throughput for local communication as $\mbox{TL}^{(1)}(M_{2}) = \sqrt{M_{2}}/(2L^2\sqrt{1+Q})$,  yielding
the achievable sum-rate:
\begin{equation*}
R_{{\rm sum}}^{(2)} (n,\alpha)= R_{{\rm c}}(\alpha) \frac{nM_{2}}{(1+Q)M_{2}^2/\mbox{TL}^{(1)}(M_{2})+n}
\end{equation*}
where TDMA is not used in this stage, as shown in Fig.~\ref{eTDMA1}.
With the optimal cluster size $M_{2}=n^{2/3}/(L^{4/3}(1+Q))$, we have $R_{{\rm sum}}^{(2)} (n,\alpha) = R_{{\rm c}}(\alpha) n^{2/3} / \left(3L^{4/3}(1+Q)\right)$.
Similarly, generalizing to a $t$-stage hierarchical protocol, we obtain:
\begin{eqnarray*}
R_{{\rm sum}}^{(t)} (n,\alpha) &=& R_{{\rm c}}(\alpha) \frac{nM_{t}}{(1+Q)M_{t}^2/\mbox{TL}^{(t-1)}(M_{t}) + n}\nonumber\\
&\stackrel{(a)}{=}&R_{{\rm c}}(\alpha)\frac{nM_{t}}{n + tL^2(\sqrt{1+Q})^{t+1}M_{t}^{(t+1)/t}}\\
&\stackrel{(b)}{=}& R_{{\rm c}}(\alpha) \frac{n^{\frac{t}{t+1}}}{(1+t)L^{\frac{2t}{t+1}}(\sqrt{1+Q})^{t}}
\end{eqnarray*}
where (a) is from Lemma~\ref{lem:NT} below
and (b) is from the optimal cluster size $M_{t} =\left( n/(L^2(\sqrt{1+Q})^{t+1})\right)^{t/(t+1)}$.

%

\begin{lemma}\label{lem:NT}  The packet throughput of local communication of stage-$t$ is given by $\mbox{TL}^{(t)}(n) =n^{\frac{t}{t+1}}/\left((t+1)L^{2}\sqrt{1+Q}^{t}\right)$.
\hfill $\QED$
\end{lemma}

\section*{Acknowledgment}

This work was supported by NSF Grant CCF 1161801.



\end{document}